\documentclass[%
 aip,
 jap,
reprint,
]{revtex4-1}

\usepackage{graphicx}
\usepackage{dcolumn}
\usepackage{bm}
\usepackage{epsfig}
\usepackage{amsmath}
\usepackage{amssymb}
\usepackage{placeins}
\usepackage{ifsym}
\usepackage{tabulary}
\usepackage{multirow}
\usepackage{booktabs}
\usepackage[dvips]{hyperref}
\hypersetup{colorlinks=true, breaklinks, urlcolor=blue, linkcolor=blue, citecolor=blue}
 
\begin{document}

\preprint{AIP/123-QED}

\title{High-intensity positron microprobe at the Thomas Jefferson National Accelerator Facility}

\author{S. Golge} 
\thanks{Author to whom correspondence should be addressed. Electronic mail: serkan.golge@nasa.gov} 
\author{B. Vlahovic} 
\affiliation{North Carolina Central University, Durham, NC 27707, USA}
\author{B. Wojtsekhowski}
\affiliation{Jefferson Laboratory, 12000 Jefferson Ave., Newport News, VA 23606, USA} 

\date{\today}

\begin{abstract}
We present a conceptual design for a novel continuous wave electron-linac based high-intensity high-brightness slow-positron production source with a projected intensity on the order of 10$^{10}$ e$^+$/s.  Reaching this intensity in our design relies on the transport of positrons (T$_+$ below 600 keV) from the electron-positron pair production converter target to a low-radiation and low-temperature area for moderation in a high-efficiency cryogenic rare gas moderator, solid Ne.  This design progressed through Monte Carlo optimizations of: electron/positron beam energies and converter target thickness, transport of the e$^+$ beam from the converter to the moderator, extraction of the e$^+$ beam from the magnetic channel, a synchronized raster system, and moderator efficiency calculations. For the extraction of e$^+$ from the magnetic channel, a magnetic field terminator plug prototype has been built and experimental results on the effectiveness of the prototype are presented.  The dissipation of the heat away from the converter target and radiation protection measures are also discussed.
\end{abstract}

\maketitle

\section{\label{sec:Introduction}Introduction}
Over the years, experts in the positron community
have recognized the need for a high-intensity and high-brightness slow positron source which outperforms available positron source intensities. Furthermore, there is great interest from the solid state and atomic physics communities in  exploring fundamental and applied research areas that are not accessible with the currently existing low-energy positron intensities.~\cite{ref:Lynn1984,Dupasquier:1985,Mills:1152001,Surko20061,Coleman2009ps}

There are many experiments that would benefit from a high-intensity high-brightness slow positron source.  For example, the 2D-Angular Correlation of Annihilated Radiation (2D-ACAR) measurement to determine fragile Fermi surface pieces of complex materials is source limited and may require several months of data accumulation.~\cite{PhysRevLett.67.382,1367-2630-14-3-035020} As demonstrated in~\cite{Mayer20101772}, the Positron Annihilation induced Auger Electron Spectroscopy (PAES)~\cite{Weiss2001363} measurement time has been significantly improved with the availability of an intense source.  Electron-positron plasma,~\cite{PhysRevLett.75.3846,surko:2333} Positronium (Ps) Bose-Einstein Condensate (BEC),~\cite{PhysRevB.49.454} Ps$_2$ molecule formation,~\cite{MillsJr2002107} and gamma-ray laser~\cite{MillsJr2004424} experiments are also among the many requiring high-intensity positron sources. 
At present, there are many table-top radioactive source-based and a few linac-based slow e$^+$ beams with intensities limited up to 10$^6$ slow e$^+$/s.~\cite{Golge:2012zz} Higher intensities have been reached at two reactor-based positron facilities: PULSTAR Reactor~\cite{Moxom2007534} and NEPOMUC Reactor~\cite{Hugenschmidt2008616} routinely operated with intensities approaching  $5 \times 10^8$ and 10$^9$ slow e$^+$/s respectively, and an electron linac-based facility, EPOS,~\cite{KrauseRehberg20063106} with a projected intensity of $5\times10^8$ slow e$^+$/s.

For some experiments, the brightness of the positron beam is more important than the intensity of the beam.~\cite{PhysRevB.31.5628,brandes:228} Thus several stages of positron (re)moderation, a process known as brightness enhancement,~\cite{Mills1980_23} are required to obtain the desired spatial resolution at the expense of losing a significant fraction of the intensity of the positrons.  Beam brightness ($\textbf{B}$) is a commonly used figure of merit to present the quality of the beam.  Here, we use the positron beam brightness definition provided in~\cite{1367-2630-14-5-055027}:
\begin{eqnarray}
\label{eq:makhov}
\textbf{B} &=& \dfrac{ Y^+_{mod} } {\theta^2 d^{+2} E{_L}}  \equiv  \dfrac{ Y^+_{mod} } { d^{+2} E_{\perp}},
\end{eqnarray}
where $d^+$ is the diameter, $\theta = \sqrt{ E_{\perp}/ E_{L}}$ is the divergence with $E_{\perp} (E_{L})$ being the transverse (longitudinal) components, and $Y^+$ is the yield of the moderated positrons. Among the low-energy high-intensity positron beam sources, the highest beam brightness has been achieved after remoderation at NEPOMUC with a value of $\sim 1.1 \times 10^7$ mm$^{-2}$ eV$^{-1}$ s$^{-1}$ and a beam spot size of $\sim$ 2 mm (FWHM) at 1 keV.~\cite{1367-2630-14-5-055027}  

In this study, we show that a 120 kW (120 MeV - 1 mA) Continuous Wave (CW) incident electron beam can provide a slow-positron intensity of up to $\sim 4.3\times 10^{10}$ e$^+$/s.  After the remoderation, we calculate that the brightness of the beam will approach $ 10^{11}$ mm$^{-2}$ eV$^{-1}$ s$^{-1}$ at 1 keV positron beam energy. There are three key elements in our design to achieve this high intensity and brightness.  First, we propose to use a high-power high-energy electron linac as it provides adjustable electron current and timing; second, we separate positrons with the desired energy range from electrons and other background radiation by employing a curved solenoid and  raster magnets, and finally we use a high-efficiency Rare Gas Moderator (RGM), such as solid Neon, which is not being used currently in linac- or reactor-based positron sources due to several challenges even though it provides the highest moderation efficiency.

Although the focus in this paper will be on the description of a low-energy positron source and beamline, it is worth mentioning that there is broad support for a positron physics program at Jefferson Lab (JLab) for high-energy physics experiments.~\cite{thomas:3}  At JLab, where an electron beam is used to probe the nucleus and the nucleons, the nuclear physics community has shown strong interest in using both polarized and unpolarized positrons for parton imaging using deeply virtual Compton scattering,~\cite{burkert:43} dispersive effects (two-photon exchange) studies in electron scattering,~\cite{arrington:13} and dark-matter searches~\cite{wojtsekhowski:149} utilizing a high-energy positron beam.  Some of these experiments would benefit from our proposed positron beamline as well since the energy distribution of the produced positrons extends up to 120 MeV with the potential to transport a several $\mu$A positron beam current at peak energies.~\cite{SerkanGolge_2010}

\section{\label{sec:Concept}The challenges and the Concept}
The experimental physics opportunities with such a low-energy positron beam are promising, but the design for an electron linac-based high-intensity source poses several challenges, including: the high initial cost of the driving electron linac, the low efficiency of positron production, the poor collection efficiency, the dissipation of deposited power in the converter used to produce positron beam, and the type of moderator. Our design addresses each of these challenges.

The proposed slow e$^+$ source will be based on an existing CW super-conducting e$^-$ linac beam at JLab. Currently, two injectors are operational at JLab and can be converted to positron production: the injectors for the Continuous Electron Beam Accelerator Facility (CEBAF) and the Free Electron Laser (FEL). For this conceptual study, we used the electron beam parameters  for the FEL injector in our computational studies as an input. The beam at FEL runs at 75 MHz (or sub-harmonics) with a micro-pulse width of 3 ps and a micro-charge of up to 13 pC (providing up to a 1 mA CW beam in a non-recovery energy regime).

Metallic film moderators (such as tungsten mesh) have conventionally been used in reactor-based and linac-based positron sources.~\cite{Golge:2012zz} The efficiency ($\eta_{++}$ = slow e$^+$/fast e$^+$) of these moderators is reported in the range of $\sim 10^{-4} -10^{-3}$.~\cite{Saito200213,Weng2004397}  On the other hand, the $\eta_{++}$ of a solid Ne\ moderator is more than an order of magnitude higher, in the range of $\sim 7\times 10^{-3}-1.4\times 10^{-2}$ with positrons emitted from a $^{22}$Na positron emitter.~\cite{Meshkov2008_5,khatri:2374} Solid Ne provides much higher efficiency than metallic moderators due to the fact that positron diffusion length (L$_+$) inside a RGM is much longer than it is inside a metallic moderator, and therefore thermalized positrons may diffuse back to the surface with higher probability before they annihilate inside the material.  Metallic moderators do provide much lower efficiency when compared with solid Ne; however; they emit slightly narrower energy spread positrons.  

Based on experimental results with a solid Ne moderator,~\cite{Meshkov2008_5,khatri:2374} using the positron kinetic energy spectrum of the $^{22}$Na as a baseline, we designed the beamline to collect and transport e$^+$ with kinetic energy (T$_+$) below 600 keV from the  e$^-$ - e$^+$  pair-production converter.  
It is important to note that the cryogenic nature of the RGM mandates that it must be positioned away from the high temperature and radiation area around the converter. Typical decay rate of the solid Ne moderator, once it has been grown, has a half-life of $\sim$ 7 days.~\cite{doi:10.1139/p96-063}  The lifetime of the solid Ne moderator is very sensitive to the changes in vacuum and temperature of the conditions in which it operates.  Therefore, transporting positrons that are suitable for moderation away from the converter area will allow us to use RGMs and evaluate other moderator options. 

\subsection{Optimized energies of the emitted positron and driving electron beams}
We divided the optimization study to find the operation energy regime into two parts. Our interest in using an RGM motivated the first part where we focused on maximizing $\eta_+$, the intensity of the emerging e$^+$ per incident e$^-$,  with positron kinetic energies below 600 keV for various converter thicknesses.
\begin{figure}[h]
\centering
\includegraphics[width=0.450\textwidth]{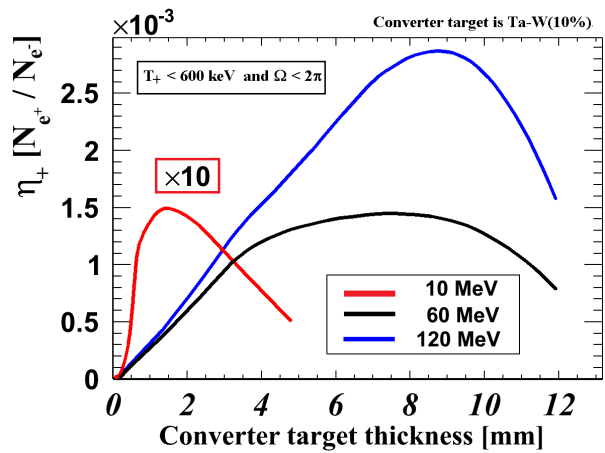}
\caption{ The yield, $\eta_+$, as a function of the converter thickness.  The yield presented includes only e$^+$ with T$_+$ $<$ 600 keV in the forward 2$\pi$. The 10-MeV curve is multiplied by 10 for visualization.}
\label{Fig_CutPositronSpectrum}
\end{figure}
\begin{figure}[h]
\centering
\includegraphics[width=0.50\textwidth]{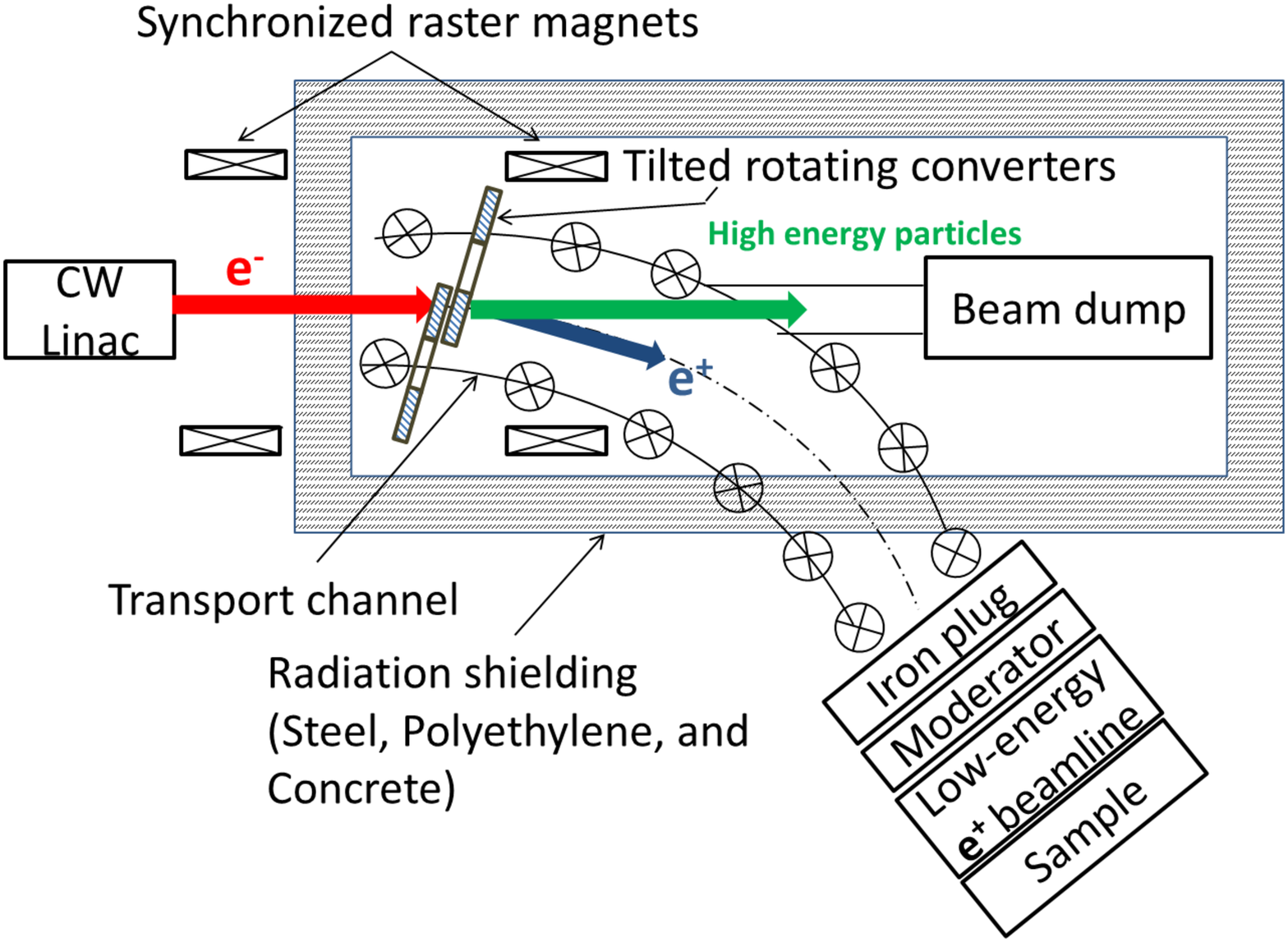}
\caption{Conceptual layout of the positron beamline. Drawing is not to scale.}
\label{Fig_SourceConcept}
\end{figure}
In the second part, we optimized the incident e$^-$ beam energy to maximize $\eta_+$  predicted by the intensity optimization.  We performed Monte Carlo simulations in the framework of the \begin{small}\textsf{GEANT4}\end{small}-based software, \begin{small}\textsf{G4beamline}\end{small}~\cite{4440461}, to evaluate these optimization parameters.

For the incident beam energy, we used 10, 60, and 120 MeV/c for various thicknesses of a  W(10\%)-Ta converter to find the highest e$^+$ yield below a kinetic energy cut of 600 keV.  In Fig.~\ref{Fig_CutPositronSpectrum}, the positron-production efficiency as a function of converter thickness is shown, where $\eta_+$ has a cut and presents only e$^+$ in the 2$\pi$ sr solid angle in the forward  direction with energies below 600 keV, per incident number of e$^-$ on the converter.  As seen in the figure, there is a broad maximum e$^+$ yield between 6 and 10 mm thicknesses with a 120 MeV/c incident electron beam, which is about a factor of two higher than the peak yield at 60 MeV/c and a factor of twenty higher than the peak yield at 10 MeV/c.  As a result, we have selected a 120 MeV/c incident beam and an 8 mm thick converter, where the efficiency is  $\eta_+ \backsim 3\times10^{-3}$ e$^+$ / incident e$^-$.  Due to the multiple scattering of charged particles in the converter, both the momentum spread and angular spread of the emitted e$^+$ are very large with $x^{\prime}=p_x/p_z$ up to $\pm$1.5 rad.  Also, the $p$ spectrum of the emerging positrons goes up to the incident beam momentum; however; only a small fraction of the positron beam can be collected and is useful. Thus, we designed the front-end capture system to collect the highest number of e$^+$ that can be transported and moderated in an RGM.  To capture a significant fraction of the emitted e$^+$ within the given energy cut, the required longitudinal field is calculated to be, $B_z=cp_\bot/\text{e}\rho \sim 2$ kG in a 6 cm inner-diameter solenoid channel, where e is the charge and $\rho$ is the Larmor radius of the particles. 
\subsection{Description of the positron source}
For the transportation of e$^+$ to the moderator we designed an arc-shaped solenoid capture and transport channel. The purpose of this curved transport channel is to transport e$^+$ away from the high-radiation and high-temperature area to be able to use a solid Ne moderator.  The photons, electrons, and positrons with high energies are much more collimated than low-energy particles and they will hit the beam dump in a straight path.  The conceptual layout of the positron source is shown in Fig.~\ref{Fig_SourceConcept}.

The curved channel has a bending radius  of 4 m and total arc length of $\sim$ 4.2 m with an arc angle of 60$^{\circ}$.  The  longitudinal field in the solenoid channel is 2 kG. The converter target is positioned inside the solenoid channel to take advantage of the full magnetic field strength when collecting positrons. Since the transport solenoid channel is curved, there will be induced drift forces on positrons.~\cite{ref:plasma_schmidt} As a result of these forces, the positrons drift away from the center.  Corrector dipole magnets, which are required to align the positron beam's central orbit offset, are super-positioned on the solenoid channel. The required integrated field along the channel is calculated to be B[G]ds[cm] = 1350 Gcm.

\begin{figure}[h]
\centering
\includegraphics[width=0.50\textwidth]{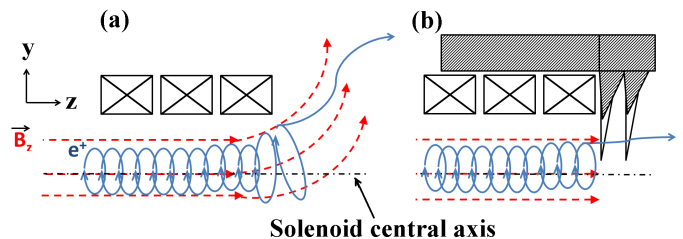}
\caption{Concept of transport through the solenoid channel (a) without and (b) with the magnetic field terminator steel plug. Solid blue lines show e$^+$ track. Dashed red lines represent magnetic field lines. Drawing is not to scale.}
\label{Fig_PlugConcept}
\end{figure}

The extraction of the positrons from the solenoid channel to a very low magnetic field area will be achieved by a magnetic field terminator plug, where the moderator will be located right after this plug.  In Fig.~\ref{Fig_PlugConcept}, the concept of the magnetic field terminator design is illustrated.
The extraction efficiency from the solenoid channel is enhanced with rapid extinction of the guide field.  Otherwise, the lowest energy, and most desirable e$^+$, will follow the diverging field lines into material surfaces and be lost.  Thus, we designed a magnetic field terminator iron plug, similar to a ``magnetic spider'' plug designed elsewhere,~\cite{Stoeffl19991} to insert at the end of the solenoid for transition to a field free area.
\begin{figure}[h]
\centering
\includegraphics[width=0.40\textwidth]{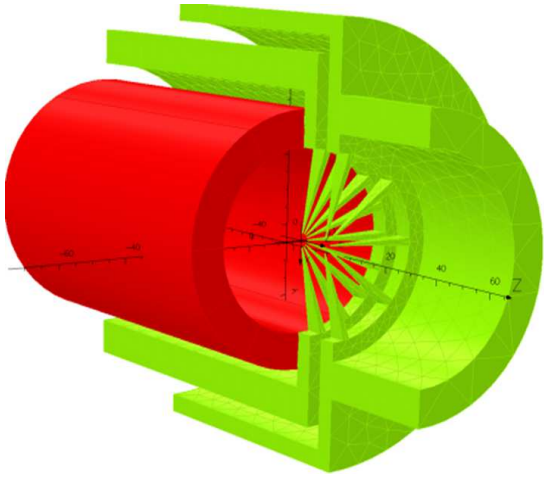}
\caption{\begin{small}\textsf{OPERA-3D (TOSCA)}\end{small} model of the magnetic plug is shown. Half of the magnetic plug is shown for detailed visualization. The red cylinder shown is the end portion of the solenoid.  The beam travels in the positive Z axis. The scale is in units of mm.}
\label{Fig_Tosca1}
\end{figure}
Figure~\ref{Fig_Tosca1} shows the end section of the solenoid (red) and half cross-section of the plug (green) designed in the framework of \begin{small}\textsf{OPERA-3D (TOSCA)}\end{small}~\cite{opera_tosca}. 

The field terminator plug is composed of two nested wedge structures. As seen in the figure, the outer jacket has a cylindrical extension. Each wedge is tapered both radially and transversely from the circular rim to the center of the plug. The wedge thickness varies from 150 $\mu$m to 15 $\mu$m at the center. The wedges, 18 in the inner structure and 10 in the outer, are separated in equal angles azimuthally. The wedges do not intersect at the center of the structure.  There is a 2 mm wide diameter hole at the center of the structure to allow the lowest energy positrons to exit the plug without interacting with the wedges.  Although the intensity of the transported positrons would slightly increase by widening the aperture; however; we optimized the aperture diameter to achieve a rapid field-termination and low amount of field leakage to the moderation area.  Further significant field reduction will be possible by using a high-permeability magnetic shield enclosure at such a low-field environment.   We designed and constructed a simpler prototype field terminator iron plug to compare magnetic field termination effectiveness against our calculations.
\subsection{Prototype magnetic field terminator plug}
\begin{figure}[h]
\centering
\includegraphics[width=0.5\textwidth]{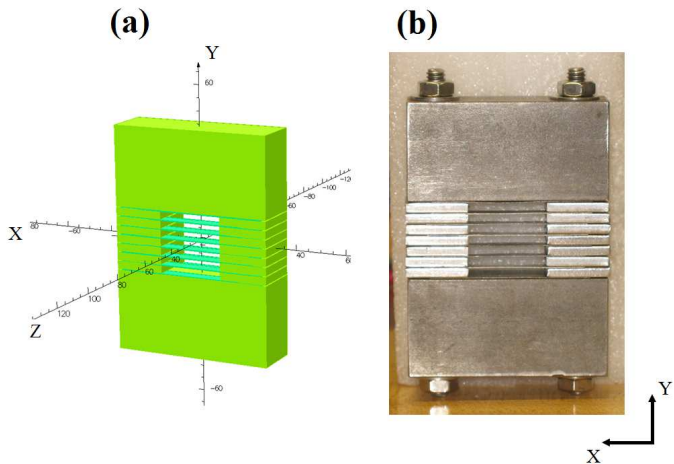}
\caption{(a) \begin{small}\textsf{OPERA-3D (TOSCA)}\end{small} simulation model and (b) manufactured prototype magnetic plug are shown. The scale in (a) is in units of mm.}
\label{Fig_ProtoPlugTosca2}
\end{figure}

We verified the effectiveness of field reduction with the manufactured prototype magnetic field terminator plug. The prototype plug was inserted in a dipole magnet with a field intensity inside of 2 kG .  In Fig.~\ref{Fig_ProtoPlugTosca2}(a), a \begin{small}\textsf{TOSCA}\end{small} model and in Fig.~\ref{Fig_ProtoPlugTosca2}(b) the manufactured prototype plug are shown.

\begin{figure}[h]
\centering
\includegraphics[width=0.45\textwidth]{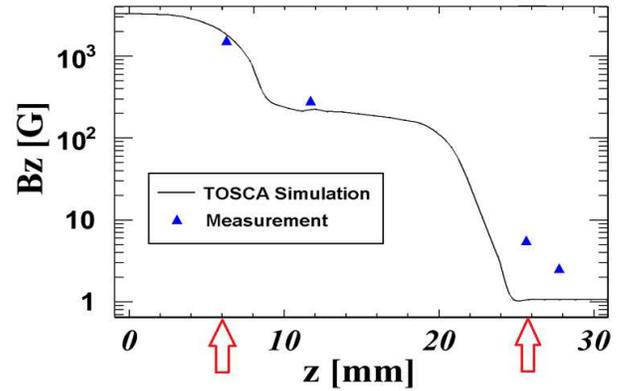}
\caption{Simulation and measurement comparison of the
prototype magnetic plug. Arrows indicate span of the plug.}
\label{Fig_ProtoResults}
\end{figure}

The large bars of the prototype plug are made of ASTM A848 grade steel ($<$ 0.01\% Carbon content), and the small fins in the middle are made of Permendur (49\% Fe, 49\%Co, and 2\% V). Both materials have very high magnetic field saturation values. In Fig.~\ref{Fig_ProtoResults},
%
%
the results of simulation and experimental measurement of the prototype magnetic plug are shown.  The red arrows show the beginning (z = 6 mm) and end (z = 25 mm) positions of the prototype plug.  As seen, the simulation and data are in good agreement; the plug reduced the field density three-fold from B$_z \sim$ 2 kG to a few Gauss.

\subsection{Simulations of the transportation of positrons through solenoid transport channel}
\begin{figure*}[]
\centering
\includegraphics[width=0.75\textwidth]{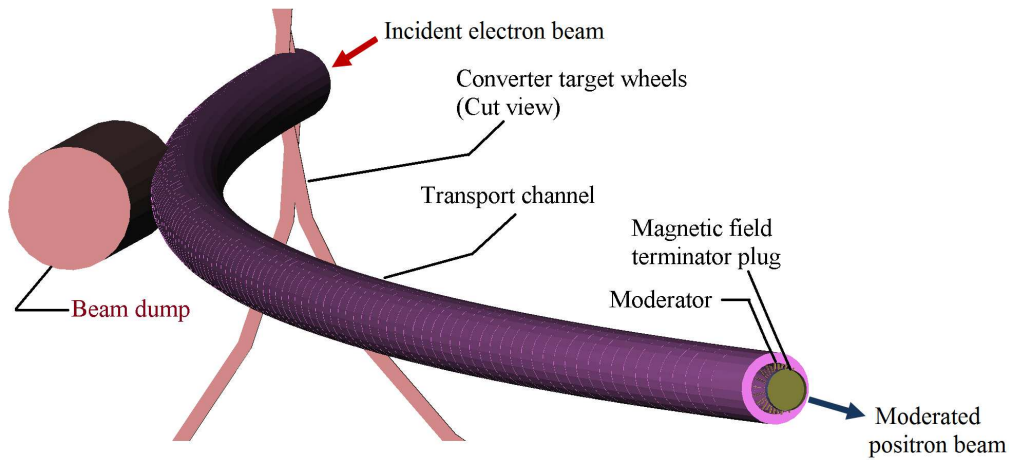}
\caption{Snapshot from the \begin{small}\textsf{G4beamline}\end{small} Monte Carlo simulation of the positron beamline. Cut view.}
\label{Fig_G4beamlineFull}
\end{figure*}

\begin{figure}[h]
\centering
\includegraphics[width=0.50\textwidth]{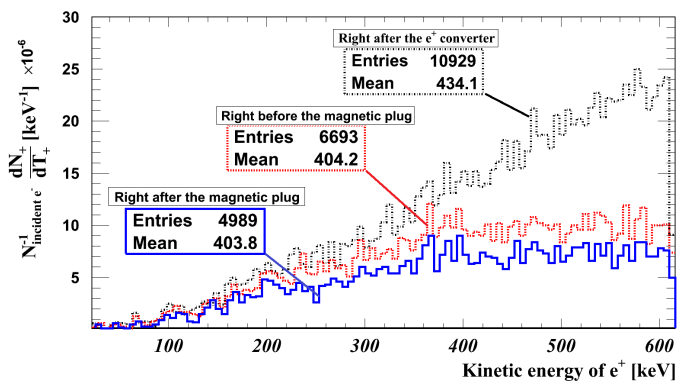}
\caption{Kinetic energy spectrums of the positrons right after the e$^+$ converter target, right before and right after the magnetic plug are shown.  Positrons shown here have a cut in energy with T$_+ < $ 600 keV.}
\label{Fig_PositronSpectrum}
\end{figure}

A snapshot from the Monte Carlo simulation of the positron beamline is shown in Fig.~\ref{Fig_G4beamlineFull}.  Here we only present the converter targets, solenoid transport channel, the magnetic plug, beam dump, and the moderator.  For the purpose of presenting a clear picture, other particles were killed during production at the converter.  In the simulation, we modeled the curved solenoid channel with hundreds of very short  straight solenoids. We verified the uniformity of the longitudinal field map inside the solenoid.  The magnetic field map of the end of the channel including the magnetic field terminator plug was imported from \begin{small}\textsf{OPERA-3D (TOSCA)}\end{small} code into the  simulation.  
In Fig.~\ref{Fig_PositronSpectrum}, positron kinetic energy spectrums from the simulation (T$_{+} < 600$ keV) right \textbf{after} the positron converter target, right \textbf{before} the magnetic field terminator plug, and right \textbf{after}  the magnetic field terminator plug are shown. From the target to the magnetic plug, approximately 60\% of the positrons are transported.  About 25\% of these positrons are lost when passing through the magnetic field terminator plug. In the end, from the positron converter to the Solid Ne moderator, about 54\% of the positrons are transported with our transport channel.

The e$^+$  transport efficiency from the e$^+$  converter to the moderator is calculated to be $\sim5\times10^{-4}$ e$^+$/incident e$^-$.  With the assumption of a 1 mA incident electron beam current, the intensity of the e$^+$ impinging on the solid Ne moderator would be $\sim 3.1\times10^{12}$ e$^+$/s within a transverse spot size of $\sigma \sim$ 8 mm (r.m.s.).  By using the reported efficiencies in the literature (0.7-1.4\%)  for solid Ne RGM, the projected slow e$^+$ intensity emitted from the solid Ne will be in the range of $2.2 - 4.3\times10^{10}$ slow e$^+$/s with T$_+ \approx$ 1-2 eV.
%
\subsection{Moderation of positrons}
During the moderation process inside materials, positrons lose energy predominantly via ionization and excitations and are rapidly thermalized within a few ps. Below a certain threshold the only possible energy loss mechanism is through phonon interaction.~\cite{PhysRevLett.57.376} After thermalization, positrons close to the surface with energies larger than the work function ($\phi_+$) may be reemitted to the surface.  The positron implantation profile in solids is well-described by a Makhovian distribution, which is given as:
\begin{eqnarray}
\label{eq:makhov}
P(z, E)&=&(m\dfrac{z^{m-1}}{z_0^{m}})\exp[-(\dfrac{z}{z_0})^{m}],
\end{eqnarray}
\begin{eqnarray}
z_{0}&=&\frac{AE^n}{\Gamma(\dfrac{1}{m}+1)}
\end{eqnarray}
where $z$ is the distance from the surface of the material, $E$ is the energy of the incident positron, $m$ is a material and energy dependent parameter, and $A$ is a fit constant.~\cite{asoka-kumar:1634} The mean penetration depth inside a material is described by $\bar z = AE^n$. Typical parameters used are n=1.6, m=2.0, and A=4.0  $\mu$g/cm$^2$.~~\cite{RevModPhys.66.841} If implanted positrons end up close to the surface, they may diffuse back to the surface before annihilation and be reemitted as slow-positrons. In insulators (e.g. solid Ne), the diffusion length is $L_+=(D_+\tau)^{1/2}$ $\sim 0.5-1\; \mu$m, in which $D_+$ is the diffusion coefficient and $\tau$ is the mean lifetime of positrons in the material.  The probability of a thermalized positron diffusing back to the surface is given by~\cite{Charlton:2005}:
\begin{eqnarray}
\label{eq:diffuse}
P_+(z) &=& \exp[-(\dfrac{z}{L_+})],
\end{eqnarray}

\begin{figure}[h]
 \centering
  \includegraphics[width=0.5\textwidth]{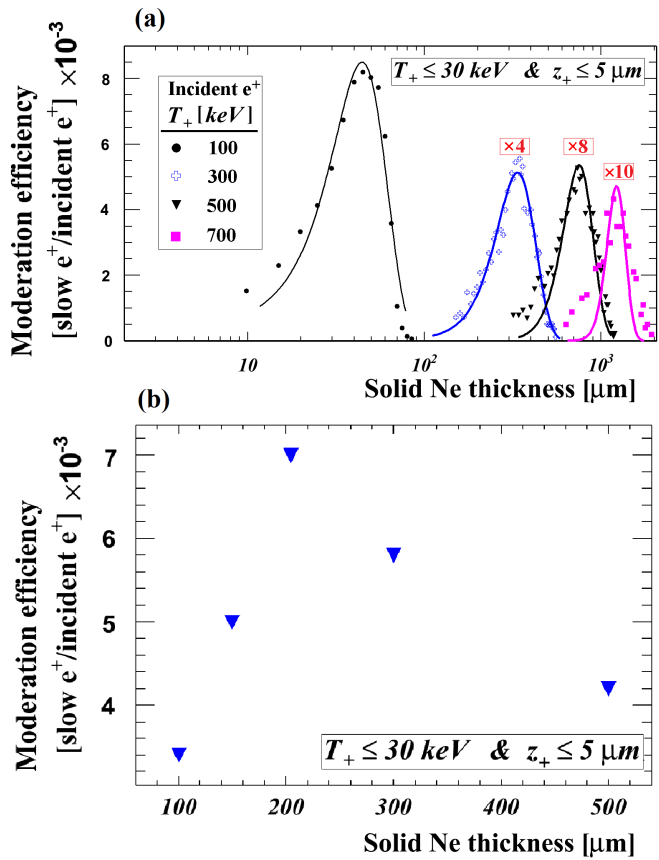}
\caption{Moderation efficiency as a function of Solid Ne thickness for (a) mono-chromatic pencil positron beam at 100, 300, 500, and 700 keV and (b) positrons right after the magnetic plug in our proposed beamline design.  To represent better, the efficiency values in (a) for 300 keV was multiplied by  a factor of four, 500 keV by eight, and 700 keV by ten.}
\label{Fig_ModerationEfficiencyT}
\end{figure}

We used \begin{small}\textsf{GEANT4}\end{small} low-energy physics data to perform Monte Carlo simulations on the implantation profile of positrons and estimated the reemission probability based on analytical and experimental results.  The probability of positron reemission in a 1-D transmission and reflection geometry is described in~\cite{ref:Fajardi} by integrating Eqs.~\ref{eq:makhov} and~\ref{eq:diffuse}.  The reemission probability for positron energies with $\bar z = 1\; \mu$m and $\bar z = 5\; \mu$m  are $\sim 35\%$ and $\sim 25\%$, respectively. With experimental data on the slow positron yield for a solid Ne moderator as provided by Mills and Gullikson (1986)~\cite{jr.:1121} as a basis, we simulated their experiment using the same cylindrical cup geometry with 8.5 mm ID and 7 mm length lined inside with a 50 $\mu$m layer of solid Ne. In  our simulation, we populated positrons using the kinetic energy spectrum of $^{22}$Na and tracked them until they arrived within 5 $\mu$m of the downstream side of the moderator surface. Based on their experimental data we assumed that positrons whose energies were reduced to 30 keV or less and within 5 $\mu$m of the surface would have an average of 30\% reemission probability weight.

The simulation results were in very good agreement with the experiment where the moderator efficiency was projected to be 0.7\%, which was the same as found by Mills and Gullikson (1986). Extending the same setup and method  for our electron-linac driven positron source in our Monte Carlo simulations, we have estimated the required thickness of the moderator to slow down fast positrons and obtain the same moderator efficiency.
As seen in Fig.~\ref{Fig_PositronSpectrum}, the average of the kinetic energy spectrum of the positrons captured by the solenoid channel is  higher when compared with the positron spectrum from a $^{22}$Na source. Thus, higher moderator thickness is required to moderate positrons with higher energies. 

By utilizing a mono-chromatic pencil positron beam, we initially made simulations to find the optimum thickness of a solid Ne moderator for the highest efficiency in transmission geometry. In Fig.~\ref{Fig_ModerationEfficiencyT}(a), simulation results of moderation efficiency for several mono-chromatic incident e$^{+}$ beams as a function of the solid Ne thickness are shown.  We used 100, 300, 500, and 700 keV e$^{+}$ beams impinging on the moderator, where again we assumed a 30\% re-emission probability for positrons whose energies were reduced below 30 keV and within the 5 $\mu$m proximity of the surface.  As expected lower energy positron beam provided higher moderation efficiency. 
When compared against the efficiency obtained with the 100 keV positron beam, the 300 keV provided about a factor of 8, the 500 keV a factor of 16, and 700 keV provided a factor of 20 less moderation efficiency.  The 100 keV beam has an optimum moderation thickness of $\sim$ 30 $\mu$m whereas, the 700 keV has $\sim$ 1050 $\mu$m.  Higher energy positrons require thicker moderators but they multiple scatter more and will be lost inside the moderator.  Therefore, it is clear that moderation becomes very inefficient for positrons with energies higher than 500-600 keV.  This study showed that the expected Solid Ne thickness for our beamline should be in the range of 100-400 $\mu$m.
 
In the latter part of this study, as a basis we used the cylindrical cup moderator geometry described in Ref.~\cite{jr.:1121}  for our beamline design simulations.  Simulation results of the optimization study to find the highest efficiency are shown in Fig.~\ref{Fig_ModerationEfficiencyT}(b).  As it can be seen from the graph, the highest moderation efficiency, which was found to be close to 0.7\%, is obtained with  $\sim$  210 $\mu$m thick solid Ne moderator.  It is worth mentioning here that the deposited power from the fast positrons in the moderator at this thickness is calculated to be approximately 50 mW.  Although this power load is much higher than that of the power load observed in radioactive source-based positron sources, we anticipate that this small amount of deposited power may be exchanged without difficulty to preserve a stable moderator temperature.  Detailed analysis of the moderator life-time and efficiency will be explored in the future.  In addition, we will explore various moderator design options such as using multiple moderators and periodically swapping them out with little beam down-time.

After the moderation, slow positrons will be extracted from the moderator by electrostatic~\cite{massoumi:1460} or magnetic~\cite{Falub2002478} focusing methods depending on the experimental needs. As it is known that the energy bandwidth of the emitted slow e$^+$ from an RGM is higher than a W moderator, the resulting beam brightness will be lower but the brightness can be significantly enhanced with brightness enhancement methods to develop a positron microprobe.  It has recently been demonstrated by Oshima $et\;al.$ (2008) that a positron beam can be brightness enhanced with one stage  remoderation through a thin transmission foil with 5$\%$ remoderation efficiency.~\cite{oshima:094916} The positron beam was produced by using a linac-driven electron beam and magnetically extracted from a solenoid channel to remoderate in a transmission remoderator. The beam spot on the sample was measured to be less than 100 $\mu$m.  This new brightness enhancement method can be directly applied to our design.  Using a single-crystal remoderator after solid Ne and taking advantage of the method utilized by Oshima $et\;al.$ (2008), we calculate that after remoderation our proposed positron source will provide beam brightness in the range of $\approx 10^{10}- 10^{11}$ mm$^{-2}$ eV$^{-1}$ s$^{-1}$ at 1 keV.

\subsection{Power dissipation in the converter}
A major challenge in a high-power linac-based positron source is the dissipation of the deposited power in the e$^{-} -$ e$^+$ production converter. As shown previously, the e$^+$ (T$_+<$ 600 keV) yield is highest with a 120 MeV e$^-$ beam incident on a 8 mm ($\sim$ 2 radiation lengths, $X_0$) thick converter.  The intensity is highest for this energy range because in the first $X_0$, the bremsstrahlung photons are produced but the electromagnetic shower reaches its maximum in the second $X_0$.  Low-energy positrons, which are of  interest, are mostly produced in the downstream $X_0$ close to the exit surface of the converter.  Therefore, it is more advantageous to use a double-layer converter over a single-layer as the power deposition would be split in the targets. The  intensity reduction with a double-layer converter exists but is negligible when compared with the single layer converter, with only $<3\%$.
\begin{figure}[h]
\centering
\includegraphics[width=0.440\textwidth]{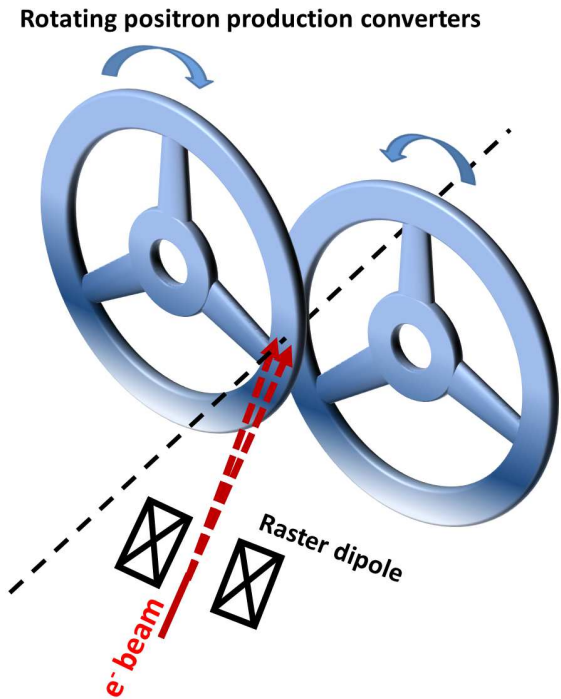}
\caption{The concept of a rotating double-layer  positron production converter target. Drawing is not to scale.}
\label{Fig_Positron_Target}
\end{figure}

When the incident e$^-$ beam power on the converter is 120 kW (120 MeV - 1 mA) , approximately 22.5\% (= 27 kW) of this incident power is deposited in a 0.8 cm thick single-layer converter that would melt the converter immediately. One well-known solution to prevent melting of the target is using a rotating double-layer wheel converter. The concept of the rotating double-layer converter is shown in Fig.~\ref{Fig_Positron_Target}. There are several efforts to realize rotating high-power targets.~\cite{Pellemoine20113,ref:Bailey2} 

In addition to the rotating converter, the incident beam in our design is rastered on the converter with magnetic steering elements (a.k.a wobbling). When the beam is rastered as the converter rotates, the effective incident electron beam size increases by orders of magnitude, thus reducing the power density and increasing the emission area for radiation cooling, the predominant mechanism at high temperatures.

In order to preserve e$^+$ beam brightness, rastering is done synchronously with two dipole sets where one dipole set is located in the upstream and the other set is located immediately downstream of the converter target to kick the positron beam back to the center of the guiding solenoid field.  We performed simulations to determine the maximum raster width that allows transportation of e$^+$ beam to the moderator without significant loss of intensity.  In these simulations, we used a converter tilted with respect to the incident beam, thereby further increasing the raster area. This study showed that with a 45$^{\circ}$ tilted converter and 1.4 cm full width raster size, about 12\% e$^+$ intensity is lost when compared with transportation without rastering.

Another key role of the raster system is sweeping low-energy electrons.  Since the solenoid captures both e$^+$ and e$^-$ from the converter, the number of e$^-$ that can reach the moderator is a factor of ten higher than the number of e$^+$.  Thus, using a raster magnet system almost completely removes the lower energy e$^-$ that would otherwise reach the moderator and deposit significant power in the cryogenic solid Ne.

\begin{figure}[h]
\centering
\includegraphics[width=0.5\textwidth]{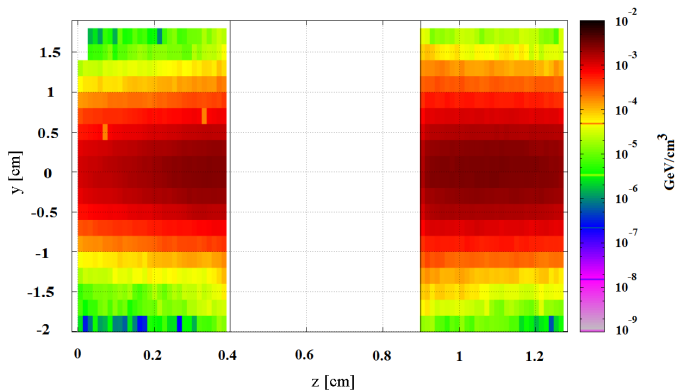}
\caption{Energy deposition density in a dual target with a total 8 mm converter thickness from a \begin{small}\textsf{FLUKA}\end{small} simulation is shown. There is a 5 mm gap between converters. The incident e$^-$ beam momentum is 120 MeV/c.}
\label{Fig_EnergyDeposit}
\end{figure}

In Fig.~\ref{Fig_EnergyDeposit}, energy deposition density in a double-layer converter is shown.

Here, each converter layer has 0.4 cm thickness with a total converter thickness of 0.8 cm. There is a 0.5 cm vacuum gap between the layers.

For the temperature calculations in a ring type rotating converter, we assumed a 45$^{\circ}$ tilted ring with a radial thickness of 1.4 cm.  As an input, we assumed an average heat capacity C$_p$ $\sim$ 170 J/kg/K,~\cite{GKWhite} the thermal conductivity of k = 100 W/m/K  around 2000 K~,\cite{Lassner_tungsten} and an average value for the emissivity of $\varepsilon$= 0.26 for tungsten.~\cite{Dmitriev1965} We targeted a conservative steady-state temperature at the interaction region, around 1800 K, which is about half the melting point of tungsten. As the target rotates, the interaction region cools down. The temperature of the interaction region in the converter drops by $\Delta$K $\sim$ 20 K in 0.5 s, in which we calculated that the target will reach a steady-state temperature with a small variation of $\Delta$K. Assuming that this $\Delta$K $\sim$ 20 K deviation is sufficient to get a uniform temperature profile, then at least a 2 Hz rotation frequency is required to reach a steady-state temperature of $\sim 1707-1845$ K at the converters. 

In Table~\ref{tab:table1}, parameters for single- and double-layer converter targets are given.  As seen in the table, the main advantage of using a double-layer converter over a single-layer is that the same target temperature goal can be reached with half the radius of a single-layer target wheel.  In addition to the deposited power calculations, we evaluated the induced eddy-current power loss due to the rotation of the wheel inside the magnetic field and calculated that the loss is insignificant due to the low speed of the converter.
%
\begin{table}[b]
\caption{\label{tab:table1}%
Comparison of parameters for single- and dual-layer rotating converter targets.}
\begin{ruledtabular}
\begin{tabular}{lccc} 
\textrm{Parameters}& \textrm{Single-Target} & \multicolumn{2}{c}{Dual-Target}\\
&&T1&T2\\
\colrule
Radius [cm] & 100 & 50 & 50 \\
Radial thickness\footnote{This is the fully rastered beam size.} [cm] & 1.4 & 1.4 &1.4 \\
Effective thickness\footnote{Since the converter is tilted, the actual thickness is smaller by a factor of sin(45$^\circ$) than the effective thickness.} [cm] & 0.8 & 0.4 & 0.4\\
Effective emission area [cm$^2$] & 1746 & 867 & 867\\
Deposited power [kW]  & 27.0 & 14.6 & 10.7 \\
Temperature [K] & 1806 &1845& 1707\\
\end{tabular}
\end{ruledtabular}
\end{table}

Pure tungsten has a higher density, and thus with less thickness the same positron intensity could be achieved. Nevertheless, it is a brittle element, which makes it challenging to work with it. A tungsten alloy, W(10\%)-Ta, is much easier to work with and more durable to the thermally induced stress loads.~\cite{ref:SLC_Feeric} Therefore, W(10\%)-Ta will be used to construct the wheel as Ta has a high-melting temperature and similar emissivity values to W as well.~\cite{Milosevic1999}
\subsection{The radiation aspect and deposited power in all elements with a high-power high-energy electron beam}
\subsubsection{The radiation aspect of the source}
\begin{figure}[h]
\centering
\includegraphics[width=0.440\textwidth]{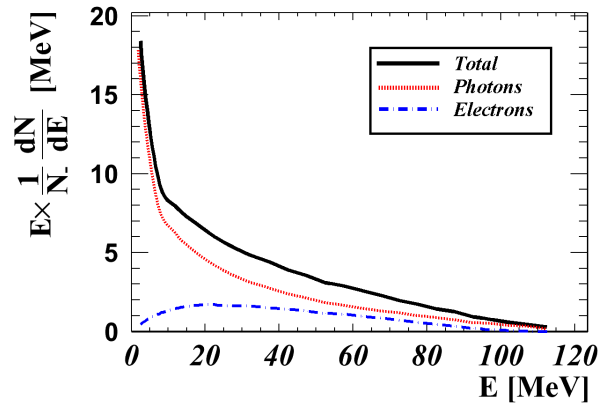}
\caption{The energy spectrum (E$\times$dN/dE) of the photons, electrons, and total of both.  The y-axis is normalized by the number of incident electrons, N$_-$.}
\label{Fig_dNdE_Secondaries.png}
\end{figure}

The major radiological concerns at an electron linac-based positron source are the emitted high energy photons, photo-neutrons, and long-lived isotopes. In Fig.~\ref{Fig_dNdE_Secondaries.png}, the energy spectra for the photons, e$^-$, and the total of both emerging from the converter in the forward direction are shown.
As is seen, a significant fraction of the power is carried out by the photons. These high-energy photons may interact with the surrounding beamline elements where, in turn, photo-neutrons may be produced, thus leading to isotope production. The attenuation of the photons can be achieved by high-Z shielding, but the shielding requirement aspects of the neutrons are dependent on the energy of the neutrons, beamline (i.e. copper vs. aluminum) materials, and shielding materials. Multi-layered radiation shielding that involves a combination of heavy and light nuclei is required to attenuate photons and neutrons.

\begin{figure}[h]
\centering
\includegraphics[width=0.450\textwidth]{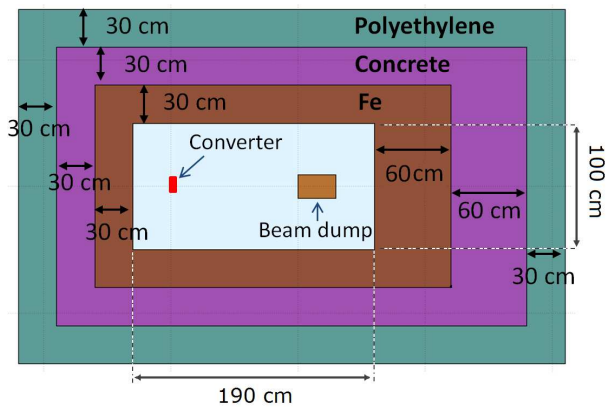}
\caption{Geometry of the model used for radiation shielding calculations in \begin{small}\textsf{FLUKA}\end{small}. In the simulation model, we only used the converter target, shielding materials, and the beam dump to calculate dose rates. In the simulations we used a copper beam dump, which has the dimensions of 15 cm in diameter and 30 cm in length as shown.}
\label{Fig_RadiationShielding}
\end{figure}

 Simulations have been performed to estimate the required thicknesses of the shielding materials for attenuation of particles and to evaluate deposited power in those elements. We calculated that at least 30 cm thick steel is required to attenuate the intensity of the photon beam to 0.1\% of its initial intensity. A local shielding configuration as shown in Fig.~\ref{Fig_RadiationShielding} was designed in the framework of \begin{small}\textsf{FLUKA}\end{small} simulation software~\cite{fluka_b,Ferrari_fluka:a}.
In this configuration, three shield walls, one upstream of the converter and two along the beamline, are constructed with 30 cm of steel followed by 30 cm of concrete and 30 cm of polyethylene. For the fourth wall, in the beam straight ahead direction downstream of the converter, the thickness of the steel and concrete is increased to 60 cm because a significant fraction of the radiation goes in the beam forward direction. A high power beam-dump will be placed along the straight line path of the incident electron beam to stop mostly collimated high-energy particles and photons (where $m/E << $ 1).  

On the other hand, typical energies of all of the emitted particles are on the order of a few MeV, and the rest of the emitted secondary particles have much larger divergence, thus they deposit their energies in the surrounding elements. Monte Carlo simulations show that this shielding configuration  reduces all types of radiation by three orders of magnitude.

At JLab, the effective radiation administrative dose limit is 250 mrem (2.5 mSv)/calendar year for radiation workers and 10 mrem (0.1 mSv)/calendar year for non-radiation workers.  According to our calculations, the shielding model we presented provides a radiation protection well below the posted limits. Although the shielding configuration attenuates the radiation significantly and prevents radiation from escaping from the source, activation would occur inside the shielding which would prevent access to the positron source for several weeks. Remote handling systems and radiation-hard materials must be used to minimize access requirements into the shielding.
\subsubsection{The deposited power in all elements}
We calculated that 25.3 kW (21 \%)of the incident 120 kW power would be deposited in a double-layer converter target, and the remaining would be deposited as: 55.7 kW (46.4 \%) in the steel plates, 30 kW (25 \%) in the beam dump, and 9 kW (7.5\%) in the solenoid. Negligible power leaves the vault from the positron beam exit port. The deposited power inside the solenoid structure was simulated using discs of coils, therefore preventing passage for the secondary beam.  We will evaluate different transport channel designs to reduce the deposited power in the solenoid, such as an open-sided magnet as described here~\cite{Coulter1994423}.
\section{Discussion and Conclusion}
Presented is the conceptual layout design of a e$^+$ source to produce a moderated slow  e$^+$ beam optimized for both incident e$^-$ beam and emitted e$^+$ beam parameters through analytical, numerical, and experimental studies.  The separation of e$^+$ from other radiation and the curved solenoid channel ending with a terminator plug coupled to a double-layered rotating converter target are distinguishing features of our positron source.  It allows us to successfully transport a majority of the created positrons from a high radiation area to a low radiation, low temperature, and low magnetic field area to facilitate the use of high-efficiency RGMs by using a linac-based high-power CW electron beam for the first time. As calculated with a 1 mA incident e$^-$ beam at 120 MeV energy on a double-layered tungsten alloy, we can transport at least $3.1\times10^{12}$  e$^+$ (T$_+ <600$ keV) to the solid Ne moderator. By using the projected moderator efficiency in the range of 0.7-1.4\% with solid Ne as cited in the literature, it is possible to produce as many as $\sim 4.3\times10^{10}$ slow e$^+$/s at the first moderation and a brightness value approaching $10^{11}$ mm$^{-2}$ eV$^{-1}$ s$^{-1}$ after remoderation.
The proposed design would provide orders of magnitude higher beam brightness, in which the anticipated performance significantly exceeds the best reported results from reactor or other available positron sources.
\begin{acknowledgements}
We would like to thank staff members of the Center for Advanced Studies of Accelerators  and Radiation Control Group at JLab,  Dr.~M.~Tiefenback, Dr.~B.~Barbiellini, and Professor ~A.~Mills for their helpful discussions and support.  This work was supported by NASA under the award NNX09AV07A and by NSF under the CREST award HRD-0833184. Authored by Jefferson Science Associates, LLC under U.S. DOE Contract No. DE-AC05-06OR23177. The U.S. Government retains a non-exclusive, paid-up, irrevocable, world-wide license to publish or reproduce this manuscript for U.S. Government purposes.
\end{acknowledgements}
\FloatBarrier

\providecommand{\noopsort}[1]{}\providecommand{\singleletter}[1]{#1}%

\end{document}